\documentstyle[12pt,epsfig]{article}
\def\bge{\begin{equation}}
\def\ene{\end{equation}}
\def\bg{\begin{eqnarray}}
\def\en{\end{eqnarray}}
\def\nn{\nonumber}

\def\S0{{\Sigma^0}}

\def\X0{{\Xi^0}}
\begin{document}
\begin{titlepage}
\title{Self-consistent description of $\Lambda$ hypernuclei\\
in the~quark-meson coupling model}
\author{
K. Tsushima$^1$~\thanks{ktsushim@physics.adelaide.edu.au}~,
K. Saito$^2$~\thanks{ksaito@nucl.phys.tohoku.ac.jp} ~and 
A. W. Thomas$^1$~\thanks{athomas@physics.adelaide.edu.au} \\
{\small $^1$Department of Physics and Mathematical Physics} \\
{\small and Special Research Center for the Subatomic Structure of Matter,} \\
{\small University of Adelaide, SA 5005, Australia} \\
{ $^2$\small Physics Division, Tohoku College of Pharmacy} \\
{\small Sendai 981, Japan} }
\maketitle
\vspace{-10cm}
\hfill ADP-97-5/T244
\vspace{10cm}
\begin{abstract}
The quark-meson coupling (QMC) model, which has been successfully used to 
describe the properties of both finite nuclei and infinite nuclear matter, 
is applied to a study of $\Lambda$ hypernuclei.
With the assumption that the (self-consistent) exchanged 
scalar, and vector, mesons  
couple only to the u and d quarks, a very weak spin-orbit force in the 
$\Lambda$-nucleus interaction is achieved automatically. 
This is a direct consequence of the 
SU(6) quark model wave function of the $\Lambda$ used
in the QMC model.
Possible implications and extensions of the present 
investigation are also discussed.
\end{abstract}
\end{titlepage}
%
In earlier work we addressed the question of whether quarks play 
an important role in finite nuclei~\cite{finite0,finite1}.
This involved quantitative investigations of the properties of
closed-shell nuclei from $^{16}$O to $^{208}$Pb.
These calculations were performed within the quark-meson coupling (QMC) model,  
first suggested by Guichon~\cite{gui}, where the interaction between 
nucleons involved the exchange of scalar ($\sigma$) and vector 
($\omega$ and $\rho$) mesons self-consistently coupled to the quarks within 
those nucleons.
Within the model it has proven possible to successfully describe not only
the properties of infinite nuclear matter~\cite{matter},   
but also the properties of finite nuclei~\cite{finite0,finite1}.
Blunden and Miller~\cite{blu}, and Jin and Jennings~\cite{jin} 
have made similar studies of the QMC model
and some phenomenological extensions.
One of the most attractive features of the QMC model is 
that it does not involve much in the way of additional complications to
Quantum Hadrodynamics (QHD)~\cite{qhd}. 
Furthermore, it produces a reasonable value for 
the nuclear incompressibility~\cite{matter}.
Recently, with an extended version of the QMC model 
which took account of the quark 
structure of the $\omega$ and $\rho$ mesons as well,
we studied the properties of finite nuclei in a unified manner, 
as well as the hadron mass changes in finite nuclei~\cite{finite2}.

Here we develop the model further to study the properties 
of $\Lambda$ hypernuclei. 
One of the main purposes of this article is to report 
the first results for 
$\Lambda$ hypernuclei calculated with a version of the QMC 
model extended to flavor SU(3),
which takes account of the quark structure of the bound $\Lambda$ 
as well as the bound nucleons. Furthermore, we treat both
the $\Lambda$ hyperon sitting outside of 
the closed-shell nuclear core, and the nucleons in the core,  
fully self-consistently, in the relativistic 
mean field approach.
Extensive results for 
$\Lambda$, $\Sigma$ and $\Xi$ hypernuclei together 
will be reported elsewhere~\cite{hyper}.

Within the Born-Oppenheimer approximation, one can derive equations of 
motion for a $\Lambda$ hypernucleus in the QMC in the same way as has been 
done for normal nuclei\cite{finite0,finite1}. Such an approach may provide 
us with important information about 
the hyperon-nucleon interaction, the deep nuclear interior and 
a possible manifestation of the quark degrees of freedom 
via the Pauli principle at the quark level.
As an example, the very weak spin-orbit interaction for 
$\Lambda$ hypernuclei, which had been 
phenomenologically suggested by Bouyssy and H\"{u}fner~\cite{bou}, 
was first explained by Brockman 
and Weise~\cite{bro} in a relativistic Hartree model,
and directly confirmed later by experiment~\cite{bru}.
However, a very strong SU(3) breaking effect was required to 
achieve this small spin-orbit force. 
An explanation in terms of quark and gluon dynamics 
was made by Pirner~\cite{pir}.
Alternatively, Noble~\cite{nob} showed that this smallness of the 
spin-orbit force could be also realized without 
any large breaking of SU(3) symmetry,  
if an $\omega \Lambda \Lambda$ tensor coupling was introduced, 
analogous to the anomalous magnetic moment of the $\Lambda$. 
However, Dover and Gal~\cite{dov} questioned whether the tensor 
coupling of the $\omega$ meson to the $\Lambda$ could 
be related to the anomalous magnetic moment,
because the spin of $\Lambda$ is entirely carried 
by the s quark in a naive SU(6) valence quark model, and this s quark  
couples exclusively to the $\phi$ meson according to the OZI rule.
Later, Jennings~\cite{jen} pointed out that within Dirac phenomenology 
the tensor coupling of the $\omega$ meson to the $\Lambda$ 
could be introduced in such a way as to guarantee that the direct $\omega$
coupling to the spin of the $\Lambda$ was zero -- as one would expect
from a simple quark model. The resulting spin-orbit force agrees with
the result obtained in the present work on the basis of an explicit
treatment of the quark structure of the $\Lambda$ moving in vector and
scalar fields that vary in space.
Many other studies of the properties of hypernuclei have been made
using relativistic, mean field theory~\cite{bro}-\cite{pro}. There has
also been a great deal of experimental 
work~\cite{bru,pov,chr,pil,aji}.

As a second example, it has also been discovered that there is an overbinding 
problem in the light $\Lambda$ hypernuclei, and 
the existence of a repulsive core or the necessity of a 
repulsive three-body force 
have been suggested to overcome the problem~\cite{bod}.
The origin of this overbinding, 
which could not be explained easily in terms of traditional nuclear 
physics, was ascribed to the Pauli principle at the quark level 
by Hungerford and Biedenharn~\cite{hun}.
Investigations of this repulsive core in the $\Lambda$-nucleus system  
have been made by Takeuchi and Shimizu, and others~\cite{tak,str}, 
based on a nonrelativistic quark model.

In the light of these earlier investigations, it seems appropriate to
investigate the (heavier) hypernuclear systems quantitatively
using a microscopic model based on quark degrees of freedom.
For this purpose, the QMC model (which is built explicitly on 
quark degrees of freedom) 
seems ideally suited, because it has already been shown to describe 
the properties of finite nuclei quantitatively.

Using the Born-Oppenheimer approximation one can derive mean-field equations 
of motion for a hypernucleus in which the quasi-particles moving 
in single-particle orbits are three-quark clusters with the quantum numbers 
of a hyperon or a nucleon. One can then construct a relativistic 
Lagrangian density at the hadronic level~\cite{finite0,finite1}, similar 
to that obtained in QHD, which produces the same equations of motion 
when expanded to the same order in $v/c$ : 
\begin{eqnarray}
{\cal L}^{HY}_{QMC} &=&  \overline{\psi}_N(\vec{r}) 
\left[ i \gamma \cdot \partial
- M_N^{\star}(\sigma) - (\, g_\omega \omega(\vec{r}) 
+ g_\rho \frac{\tau^N_3}{2} b(\vec{r}) 
+ \frac{e}{2} (1+\tau^N_3) A(\vec{r}) \,) \gamma_0 
\right] \psi_N(\vec{r}) \nn
\\
&+& \overline{\psi}_Y(\vec{r}) 
\left[ i \gamma \cdot \partial
- M_Y^{\star}(\sigma)
- (\, g^Y_\omega \omega(\vec{r}) 
+ g^Y_\rho I^Y_3 b(\vec{r}) 
+ e Q_Y A(\vec{r}) \,) \gamma_0 
\right] \psi_Y(\vec{r}) \nn \\
  &-& \frac{1}{2}[ (\nabla \sigma(\vec{r}))^2 +
m_{\sigma}^2 \sigma(\vec{r})^2 ]
+ \frac{1}{2}[ (\nabla \omega(\vec{r}))^2 + m_{\omega}^2
\omega(\vec{r})^2 ] \nn \\
 &+& \frac{1}{2}[ (\nabla b(\vec{r}))^2 + m_{\rho}^2 b(\vec{r})^2 ]
+ \frac{1}{2} (\nabla A(\vec{r}))^2, 
\label{lag}
\end{eqnarray}
where $\psi_N(\vec{r})$ ($\psi_Y(\vec{r})$)  
and $b(\vec{r})$ are respectively the
nucleon (hyperon) and the $\rho$ 
meson (the time component in the third direction of
isospin) fields, while $m_\sigma$, $m_\omega$ and $m_{\rho}$ are 
the masses of the $\sigma$, $\omega$ and $\rho$ mesons.
$g_\omega$ and $g_{\rho}$ are the $\omega$-N and $\rho$-N
coupling constants which are related to the corresponding 
(u,d)-quark-$\omega$, $g_\omega^q$, and 
(u,d)-quark-$\rho$, $g_\rho^q$, coupling constants as
$g_\omega = 3 g_\omega^q$ and 
$g_\rho = g_\rho^q$~\cite{finite0,finite1}.


In an approximation where the $\sigma$, $\omega$ and $\rho$ mesons couple
only to the u and d quarks (ideal mixing of the $\omega$ and $\phi$ 
mesons, and the OZI rule are assumed),
the coupling constants in the hyperon sector
are obtained as $g^Y_\omega = (n_0/3) g_\omega$, and 
$g^Y_\rho = g_\rho = g_\rho^q$, with $n_0$ being the number of
the valence u and d quarks in the hyperon Y. $I^Y_3$ and $Q_Y$
are the third component of the hyperon isospin and its 
charge, respectively. The field dependent $\sigma$-N and $\sigma$-Y
coupling strengths predicted by the QMC model,  
$g_\sigma(\sigma)$ and  $g^Y_\sigma(\sigma)$, 
are defined by,
\bg
M_N^{\star}(\sigma) &\equiv& M_N - g_\sigma(\sigma)
\sigma(\vec{r}) ,  \\
M_Y^{\star}(\sigma) &\equiv& M_Y - g^Y_\sigma(\sigma)
\sigma(\vec{r}) , \label{coupny}
\en
where $M_N$ ($M_Y$) is the free nucleon (hyperon) mass.
Note that the dependence of these coupling constants on the applied
scalar field must be calculated self-consistently within the quark
model. Hence, unlike QHD, even though $g^Y_\sigma / g_\sigma$ may be
$\frac{2}{3}$ in free space, this will not necessarily  be the case in
nuclear matter.
More explicit expressions for $g^Y_\sigma(\sigma)$ 
and $g_\sigma(\sigma)$ will be given later. From 
the Lagrangian density 
in Eq.~(\ref{lag}), one gets a set of 
equations of motion for the hypernuclear system, 
\begin{eqnarray}
& &[i\gamma \cdot \partial -M^{\star}_N(\sigma)-
(\, g_\omega \omega(\vec{r}) + g_\rho \frac{\tau^N_3}{2} b(\vec{r}) 
 + \frac{e}{2} (1+\tau^N_3) A(\vec{r}) \,) 
\gamma_0 ] \psi_N(\vec{r}) = 0, \label{eqdiracn}\\
& &[i\gamma \cdot \partial - M^{\star}_Y(\sigma)-
(\, g^Y_\omega \omega(\vec{r}) + g_\rho I^Y_3 b(\vec{r}) 
+ e Q_Y A(\vec{r}) \,) 
\gamma_0 ] \psi_Y(\vec{r}) = 0, \label{eqdiracy}\\
& &(-\nabla^2_r+m^2_\sigma)\sigma(\vec{r}) = 
- [\frac{\partial M_N^{\star}(\sigma)}{\partial \sigma}]\rho_s(\vec{r})  
- [\frac{\partial M_Y^{\star}(\sigma)}{\partial \sigma}]\rho^Y_s(\vec{r}),
\nn \\
& & \hspace{7.5em} \equiv g_\sigma C_N(\sigma) \rho_s(\vec{r})
    + g^Y_\sigma C_Y(\sigma) \rho^Y_s(\vec{r}) , \label{eqsigma}\\
& &(-\nabla^2_r+m^2_\omega) \omega(\vec{r}) =
g_\omega \rho_B(\vec{r}) + g^Y_\omega 
\rho^Y_B(\vec{r}) ,\label{eqomega}\\
& &(-\nabla^2_r+m^2_\rho) b(\vec{r}) =
\frac{g_\rho}{2}\rho_3(\vec{r}) + g^Y_\rho I^Y_3 \rho^Y_B(\vec{r}),  
 \label{eqrho}\\
& &(-\nabla^2_r) A(\vec{r}) = 
e \rho_p(\vec{r}) 
+ e Q_Y \rho^Y_B(\vec{r}) ,\label{eqcoulomb}
\end{eqnarray}
where, $\rho_s(\vec{r})$ ($\rho^Y_s(\vec{r})$), $\rho_B(\vec{r})$ 
($\rho^Y_B(\vec{r})$), $\rho_3(\vec{r})$ and 
$\rho_p(\vec{r})$ are the scalar, baryon, third component of isovector,   
and proton densities at the position $\vec{r}$ in 
the hypernucleus~\cite{finite1,finite2}.    
On the right hand side of Eq.~(\ref{eqsigma}),
a new, and characteristic feature of QMC beyond QHD~\cite{qhd,ruf,chr}
appears, namely,
$- \frac{\partial M_N^{\star}(\sigma)}{\partial \sigma} = 
g_\sigma C_N(\sigma)$ and 
$- \frac{\partial M_Y^{\star}(\sigma)}{\partial \sigma} = 
g^Y_\sigma C_Y(\sigma)$, where $g_\sigma \equiv g_\sigma (\sigma=0)$ and 
$g^Y_\sigma \equiv g^Y_\sigma (\sigma=0)$. 
The effective mass for the hyperon Y is defined by,
\begin{equation}
\frac{\partial M_Y^{\star}(\sigma)}{\partial \sigma}
= - n_0 g_{\sigma}^q \int_{bag} d\vec{y} 
\ {\overline \psi}_{u,d}(\vec{y}) \psi_{u,d}(\vec{y})
\equiv - n_0 g_{\sigma}^q S_Y(\sigma) = - \frac{\partial}{\partial \sigma}
\left[ g^Y_\sigma(\sigma) \sigma \right],
\end{equation}
with the MIT bag model quantities~\cite{finite0,finite1}, 
\begin{eqnarray}
& &M_Y^{\star}(\sigma) =
\frac{n_0\Omega^{\star}(\sigma) 
+ (3-n_0)\Omega^{\star}_s - z_Y}{R_Y^{\star}}
+ \frac{4}{3}\pi ({R_Y^{\star}})^3 B ,\nn \\
& &S_Y(\sigma) = \frac{\Omega^{\star}(\sigma)/2 
+ m_{u,d}^{\star}(\sigma)R_Y^{\star}(\Omega^{\star}(\sigma)-1)}
{\Omega^{\star}(\sigma)(\Omega^{\star}(\sigma)-1) 
+ m_{u,d}^{\star}(\sigma)R_Y^{\star}/2},\nn \\ 
& &
\Omega^{\star}(\sigma) = \sqrt{x^2 + (R_Y^{\star}m_{u,d}^{\star}(\sigma))^2}, 
\quad \Omega_s^{\star} = \sqrt{x_s^2 + (R_Y^{\star}m_s)^2},\quad 
m_{u,d}^{\star}(\sigma) = m_{u,d} - g_{\sigma}^q \sigma (\vec{r}), \nn \\
& &C_Y(\sigma) = S_Y(\sigma)/S_Y(0), \quad
g^Y_{\sigma} \equiv n_0 g_{\sigma}^q S_Y(0) 
= \frac{n_0}{3} g_\sigma S_Y(0)/S_N(0) 
\equiv \frac{n_0}{3}g_\sigma \Gamma_{Y/N}. \label{mit}
\end{eqnarray}
Here, $z_Y$, $B$, $x$, $x_s$ and $m_{u,d,s}$ are the parameters 
for the sum of the c.m.
and gluon fluctuation effects,
bag pressure, lowest eigenvalues for the (u, d) and s quarks, respectively, 
and the corresponding current quark masses with $m_u = m_d$.
$z_N$ and $B$ ($z_Y$) are fixed by fitting the nucleon (hyperon) mass 
in free space. 

The bag radii in-medium, $R_{N,Y}^\star$, are obtained
by the equilibrium condition\\ 
$d M_{N,Y}^{\star}(\sigma)/{d R_{N,Y}}|_{R_{N,Y}=R_{N,Y}^{\star}} = 0$.
The bag parameters calculated and chosen for the present study are,
$(z_N,\, z_\Lambda) = (3.295,\, 3.131)$, 
$(R_N,\, R_\Lambda) = (0.800,\, 0.806)$ fm (in free space),
$B^{1/4} = 170$ MeV, $(m_u,\, m_d,\, m_s) = (5,\, 5,\, 250)$ MeV. 
The parameters associated with the u and d quarks are those found in 
our previous investigations~\cite{finite1}. 
The value for the mass of the s quark in the MIT bag was 
chosen to be $279$ MeV, in order to reproduce the mass of the $\Lambda$
in that model~\cite{mitbag}. However, the final results 
turn out to be insensitive 
to the values of many of these parameters~\cite{finite0,finite1}.
The value for $\Gamma_{Y/N}$ turned out to be almost unity for all hyperons, 
so we can use $\Gamma_{Y/N} = 1$ in practice~\cite{finite2}.
At the hadron level, the entire information on the quark dynamics is condensed
in $C_{N,Y}(\sigma)$ of Eq.~(\ref{eqsigma}).
Furthermore, when this $C_{N,Y}(\sigma) = 1$, which corresponds to 
a structureless nucleon or hyperon, the equations of motion
given by Eqs.~(\ref{eqdiracn})-(\ref{eqcoulomb})
can be identified with those derived from QHD~\cite{ruf,mar,chi}, 
except for the terms arising from the tensor coupling and the non-linear 
scalar field interaction introduced beyond naive QHD.
The parameters at the hadron level, which are already fixed by the study of 
infinite nuclear matter and finite nuclei~\cite{finite1}, 
are as follows: $m_\omega = 783$ MeV, $m_\rho = 770$ MeV, $m_\sigma = 418$ MeV, 
$e^2/4\pi = 1/137.036$, $g^2_\sigma/4\pi = 3.12$, $g^2_\omega/4\pi = 5.31$ 
and $g^2_\rho/4\pi = 6.93$. 

For practical calculations, it has been found that the scalar 
densities $C_{N,Y}(\sigma)$ can be parametrized as a linear 
form in the $\sigma$ field~\cite{finite0,finite1,finite2},
\begin{equation}
C_{N,Y} (\sigma) = 1 - a_{N,Y} \times (g_{\sigma} \sigma(\vec{r})) ,
\label{cysigma}
\end{equation}
where $g_{\sigma} \sigma = (3 g^q_\sigma S_N(0)) \sigma$ in MeV, 
and the value for the $\Lambda$ hyperon is, 
$a_\Lambda = 9.25 \times 10^{-4}$ 
(MeV$^{-1}$) for $m_u$ = 5 MeV, $m_s$ = 250 MeV, $R_N$ = 0.8 fm 
$\simeq R_\Lambda$. This parametrization works very well up to 
about three times normal nuclear density, $\rho_B \simeq 3 \rho_0$, 
with $\rho_0 \simeq 0.15$ fm$^{-3}$.
Using this parametrization, one can write down the explicit 
expression for the effective mass of the hyperon $Y$:
\begin{equation}
M^{\star}_Y(\sigma) \equiv M_Y - g^Y_\sigma(\sigma) \sigma(\vec{r})  
\simeq M_Y - \frac{n_0}{3} 
g_\sigma \left[ 1 - \frac{a_Y}{2} (g_\sigma 
\sigma(\vec{r})) \right] \sigma(\vec{r}).
\label{mass} 
\end{equation}

The origin of the spin orbit force for a composite nucleon moving
through scalar and vector fields which vary with position was explained
in detail in Ref.\cite{finite0} -- c.f. sect. 3.2. The situation for the
$\Lambda$ is different in that, in an 
SU(6) quark model, the u and d quarks are coupled 
to spin zero, so that  
the spin of the $\Lambda$ is carried by the s quark. 
As the $\sigma$-meson is viewed here as a convenient parametrization of 
two-pion-exchange and the $\omega$ and $\rho$ are non-strange, it seems 
reasonable to assume that  
the $\sigma$, $\omega$ and $\rho$ mesons couple only to 
the u and d quarks. The direct contributions to the spin-orbit interaction 
from these mesons (derived in sect. 3 of Ref.\cite{finite0}) 
then vanish due to the flavor-spin structure.
Thus, the spin-orbit interaction, 
$V^\Lambda_{S.O.}(\vec{r}) \vec{l}\cdot\vec{s}$, 
arises entirely from Thomas precession:  
\begin{equation}
V^\Lambda_{S.O.}(\vec{r}) \vec{l}\cdot\vec{s} 
= - \frac{1}{2} {\vec{v}_\Lambda} \times 
\frac{d \vec{v_\Lambda}}{dt} \cdot \vec{s}
= - \frac{1}{2 M^{\star 2}_\Lambda (\vec{r}) r}
\, \left( \frac{d}{dr} [ M^\star_\Lambda (\vec{r}) 
+ g^\Lambda_\omega \omega(\vec{r}) ] \right) \vec{l}\cdot\vec{s} ,
\label{so}
\end{equation}
where, $\vec{v}_\Lambda = \vec{p}_\Lambda/M^\star_\Lambda$, 
is the velocity of the $\Lambda$ in the rest frame of the $\Lambda$
hypernucleus, and the acceleration,
$d\vec{v}_\Lambda/dt$, is obtained from the Hamilton equations of motion  
applied to the leading order Hamiltonian of the QMC model~\cite{finite0}.
Because the contributions from the effective mass of the $\Lambda$, 
$M^\star_\Lambda (\vec{r})$, 
and the vector potential, $g^\Lambda_\omega \omega(\vec{r})$, are 
approximately equal and opposite
we quite naturally expect a very 
small spin-orbit interaction for the $\Lambda$ in the hypernucleus. 
Although the spin-orbit splittings for the nucleon calculated 
in QMC are already somewhat smaller than those calculated 
in QHD~\cite{finite1}, 
we can expect even much smaller spin-orbit splittings for the $\Lambda$ 
in QMC, by comparing the expressions obtained for the 
spin-orbit potentials between the 
nucleon and $\Lambda$~\cite{finite0}. (See also Fig.~\ref{spectra}.)
In order to include the spin-orbit potential of Eq.~(\ref{so}) correctly,  
we added perturbatively the correction due to the vector potential, 
$ -\frac{2}{2 M^{\star 2}_\Lambda (\vec{r}) r}
\, \left( \frac{d}{dr} g^\Lambda_\omega \omega(\vec{r}) \right) 
\vec{l}\cdot\vec{s}$, 
to the single-particle energies obtained with the Dirac 
equation, Eq.~(\ref{eqdiracy}). This is necessary because the Dirac 
equation corresponding to Eq.~(\ref{lag})
leads to a spin-orbit force which does not correspond to the underlying
quark model, namely:
\begin{equation}
V^\Lambda_{S.O.}(\vec{r}) \vec{l}\cdot\vec{s} 
= - \frac{1}{2 M^{\star 2}_\Lambda (\vec{r}) r}
\, \left( \frac{d}{dr} [ M^\star_\Lambda (\vec{r}) 
- g^\Lambda_\omega \omega(\vec{r}) ] \right) \vec{l}\cdot\vec{s} .
\label{sodirac}
\end{equation}
This correction to the spin-orbit force, which appears naturally in the
QMC model, may also be modelled at the hadronic level of the Dirac equation by 
adding a tensor interaction.

We are now in a position to discuss the results.
As already mentioned, the calculations are fully self-consistent 
for all fields appearing 
in Eqs.~(\ref{eqdiracn})-(\ref{eqcoulomb}). 
In principle, the existence of the $\Lambda$ outside of the nuclear core 
breaks spherical symmetry, and one should include this in a truly rigorous
treatment. We neglect this effect, since it is expected to be 
of little importance for spectroscopic calculations~\cite{fur,coh}. 
However, we do include the response of the nuclear core 
arising from the self-consistent 
calculation, which is significant for a description of the baryon currents and
magnetic moments, and a pure relativistic 
effect~\cite{coh,coh2}. We will discuss later the 
self-consistency effects between the nuclear core and the $\Lambda$. 

\begin{figure}[hbt]
\begin{center}
\epsfig{file=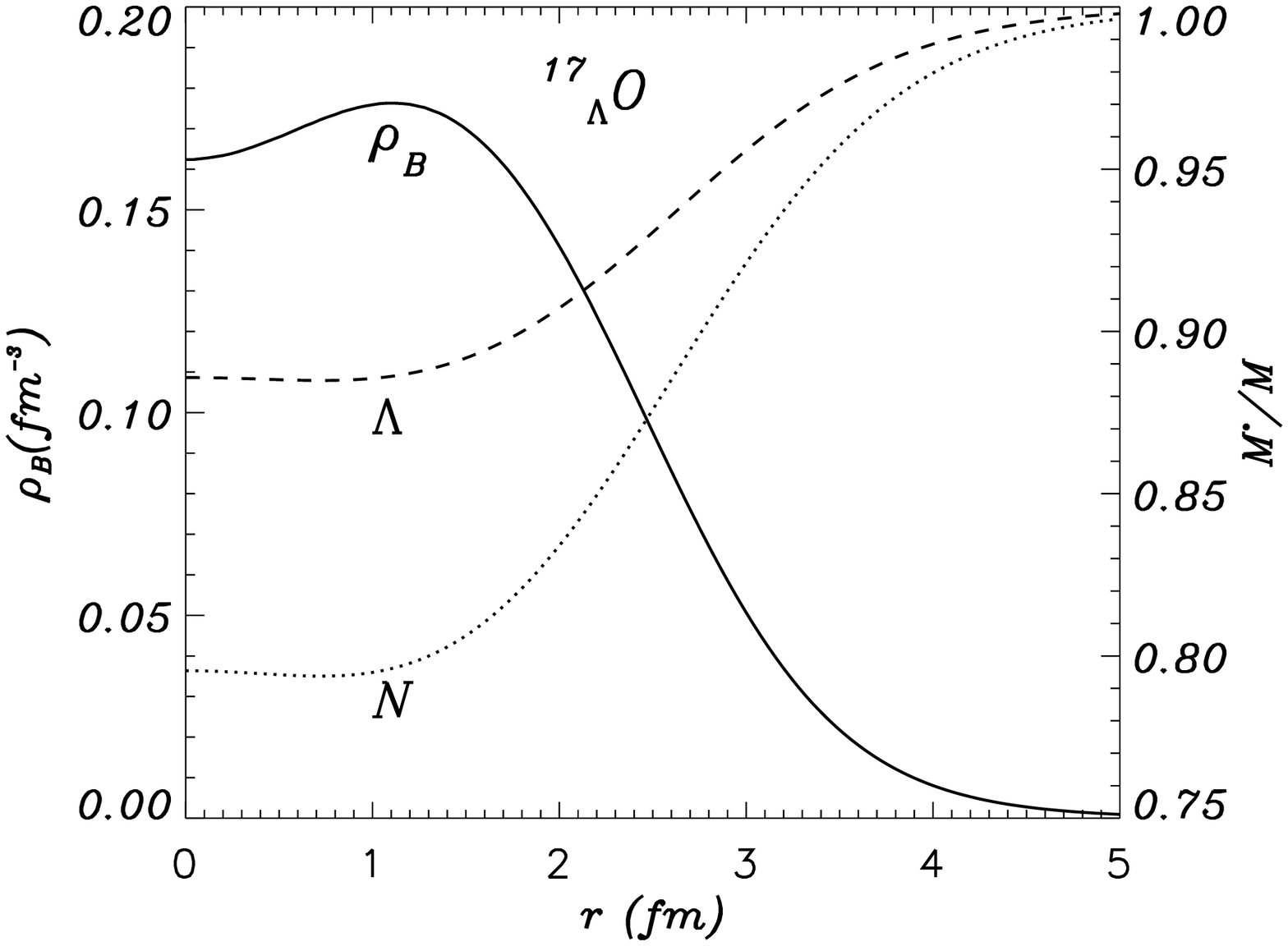,height=7cm}
\caption{(a)}
\label{ymasso}
\end{center}
\end{figure}

In Fig.~\ref{ymasspb}, we show the effective masses of 
the nucleon and $\Lambda$ 
as well as the baryon densities calculated  
for (a): $^{17}_\Lambda$O, (b): $^{41}_\Lambda$Ca and (c): $^{209}_\Lambda$Pb.
The results are for the $1s_{1/2}$ $\Lambda$ state, 
where effects of the $\Lambda$ on the whole system 
are expected to be the largest.
The effective masses in the hypernuclei, $^{17}_\Lambda$O, 
$^{41}_\Lambda$Ca and $^{209}_\Lambda$Pb 
behave in a similar manner as the distance 
$r$ from the center of each nucleus increases (the baryon density decreases).
The decrease of the baryon densities around the center in 
$^{17}_\Lambda$O in Fig.~\ref{ymasso}(a), which one might think a  
shortcoming of the present treatment, is also expected in  
nonrelativistic potential model calculation~\cite{neg}.

The calculated $\Lambda$ single-particle energies for the 
closed-shell nuclear core plus one $\Lambda$ configuration 
are listed in Table~\ref{spenergy}, with the experimental data~\cite{chr,aji}. 
At a glance, one can easily see that the spin-orbit splittings in the 
present calculation are very small for all hypernuclei.
These small spin-orbit splittings, tend to be smaller as the baryon 
density increases, 
or the atomic number increases. This smallness of the spin-orbit splitting 
is a very promising achievement in the present approach.
Concerning the single-particle energy levels, although a direct comparison 
with the data is not precise due to the
different configurations, the calculated results seem to lead to 
overestimates. In order to make an estimate of 
the difference due to this different configuration, we calculated 
the following quantity. By removing one $1p_{3/2}$ neutron in 
$^{16}$O, and putting a $\Lambda$ as experimentally observed, 
we calculated in the same way as for $^{17}_\Lambda$O-which 
means the nuclear core was still treated as spherical.
In this case the calculated energy for the $1s_{1/2}$ $\Lambda$ is -19.9 MeV, 
to be compared with the value -20.5 MeV of the present calculation.
For larger atomic numbers, the difference is smaller. 
(See also Table~\ref{spenergy}.)
We also attempted the calculation using the scaled coupling constant,   
$0.93 \times g_\sigma^\Lambda (\sigma =0)$, which reproduces the  
empirical single-particle energy for the $1s_{1/2}$ in 
$^{41}_\Lambda$Ca, -20.0 MeV~\cite{chr}.
The results obtained using this scaled coupling constant 
are also listed in Table 1, denoted by $^{41}_\Lambda$Ca$^\star$
and $^{209}_\Lambda$Pb$^\star$. 
Then, one can easily understand that the QMC model does not 
require a large SU(6) (SU(3))
breaking effect in this respect (7 \%) to reproduce the 
empirical single-particle energies.  
Furthermore, the model still achieves 
the very small spin-orbit interaction for $\Lambda$ in hypernuclei, 
without introducing the tensor coupling of the $\omega$ meson, 
which may be contrasted with the QHD type calculations~\cite{chi}.
We note that the spin-orbit splittings for the $\Lambda$ (or hyperon)  
single-particle energies in hypernuclei are not well determined 
by the experiments~\cite{pov}-\cite{aji}. 
%
%
\begin{table}[htbp]
\begin{center}
\caption{
$\Lambda$ single-particle energies (in MeV)
for closed-shell nuclear core plus one $\Lambda$
configuration. The values for $^{41}_\Lambda$Ca$^\star$ and 
$^{209}_\Lambda$Pb$^\star$ are calculated using the 
scaled coupling constant $0.93 \times g_\sigma^\Lambda (\sigma = 0)$,  
which reproduces  
the empirical single particle-energy for the $1s_{1/2}$ in 
$^{41}_\Lambda$Ca, -20.0 MeV~\protect\cite{chr}.
For reference, we list the experimental data for
$^{16}_\Lambda$O and $^{40}_\Lambda$Ca of Ref.~\protect\cite{chr} 
denoted by $a$,
and for $^{89}_\Lambda$Y  and $^{208}_\Lambda$Pb of
Ref.~\protect\cite{aji} denoted by $b$, respectively. 
Spin-orbit splittings are not well determined by the experiments. 
($^*$ fit)}
\label{spenergy}
\begin{tabular}[t]{c|ccccccccccc}
\hline \\
&$^{17}_\Lambda$O&$^{16}_\Lambda$O 
&$^{41}_\Lambda$Ca&$^{41}_\Lambda$Ca$^\star$&$^{40}_\Lambda$Ca
&$^{49}_\Lambda$Ca
&$^{91}_\Lambda$Zr&$^{89}_\Lambda$Y
&$^{209}_\Lambda$Pb&$^{209}_\Lambda$Pb$^\star$&$^{208}_\Lambda$Pb\\ \\
\hline \\
$1s_{1/2}$
&-20.5 &-12.5$^a$&-27.7 &-20.0$^*$&-20.0$^a$&-29.3  
&-32.8 &-22.5$^b$&-35.9 &-27.4    &-27.0$^b$\\
$1p_{3/2}$& -9.2 & &-18.9 &-12.6 & &-20.8 &-26.4 & &-31.9 &-23.8 & \\
$1p_{1/2}$& -9.1 &-2.5$^a$&-18.8 &-12.5 &-12.0$^a$&-20.8
&-26.4 &-16.0$^b$&-31.9 &-23.8 &-22.0$^b$\\
$1d_{5/2}$& &($1p$) & -9.5 &-4.7 &($1p$) &-11.8 &-19.3 &($1p$) &-27.1 
&-19.5 &($1p$) \\
$2s_{1/2}$& & & -8.0 &-3.6 & &-10.3 &-17.4 & &-25.4 &-17.9 & \\
$1d_{3/2}$& & & -9.4 &-4.7 & &-11.8 &-19.2 &-9.0$^b$&-27.1 
&-19.5 &-17.0$^b$\\
$1f_{7/2}$& & &   &  &       & -2.8 &-11.6 &($1d$) &-21.8 &-14.7&($1d$) \\
$2p_{3/2}$& & &   &  & &      & -9.4 &            &-19.4 &-12.6& \\
$1f_{5/2}$& & &   &  & &      &-11.5 &-2.0$^b$&-21.7 &-14.6 
&-12.0$^b$\\
$2p_{1/2}$& & &   &  & &      & -9.3 &($1f$)        &-19.4 &-12.6 &($1f$) \\
$1g_{9/2}$& & &   &  & &      & -3.7 &            &-16.0 &-9.5 & \\
$1g_{7/2}$& & &   &  & &      &      &            &-15.9 &-9.4 &-7.0$^b$\\
$1h_{11/2}$& & & & & & & & &-9.8 &-4.0 &($1g$) \\
$2d_{5/2}$& & & & & & & & &-13.2 &-7.1 & \\
$2d_{3/2}$& & & & & & & & &-13.2 &-7.1 & \\
$1h_{9/2}$& & & & & & & & & -9.7 &-3.9 & \\
$3s_{1/2}$& & & & & & & & &-12.1 &-6.2 & \\
$2f_{7/2}$& & & & & & & & & -6.9 &-1.8 & \\
$3p_{3/2}$& & & & & & & & & -5.6 &-1.1 & \\
$2f_{5/2}$& & & & & & & & & -6.8 &-1.7 & \\
$3p_{1/2}$& & & & & & & & & -5.6 &-1.1 & \\
$1i_{13/2}$& & & & & & & & &-3.4 &---  & \\
\end{tabular}
\end{center}
\end{table}
%
%

In Table~\ref{rms}, we list the calculated binding energy per baryon, $-E/A$, 
rms charge radius, $r_{ch}$, and rms radii of the $\Lambda$ and 
the neutron and proton distributions ($r_\Lambda$, $r_n$ and $r_p$, 
respectively), for the $1s_{1/2}$ and $1p_{3/2}$
$\Lambda$ configurations. The rms charge radius is calculated by convolution
with a proton form factor~\cite{finite1}.
For comparison, we also give these quantities without 
a $\Lambda$-i.e., for normal finite nuclei. The differences in values 
for finite nuclei and hypernuclei listed in Table~\ref{rms} 
reflect the effects of the $\Lambda$, 
through the self-consistency procedure. One can easily see that the effects of 
the $\Lambda$ become weaker as the atomic number becomes larger,  
and the $\Lambda$ binding energy becomes smaller.

%
%
\begin{table}[htbp]
\begin{center}
\caption{Binding energy per baryon, $-E/A$ (in MeV), rms charge radius, 
$r_{ch}$, and rms radii of the $\Lambda$, $r_\Lambda$, 
neutron, $r_n$, and proton, $r_p$ (in fm). ($^*$ fit)
}
\label{rms}
\begin{tabular}[t]{ccccccc}
\\
\hline
 &$\Lambda$ state& $-E/A$ & $r_{ch}$ & $r_\Lambda$ & $r_n$ & $r_p$ \\ 
\hline 
$^{17}_\Lambda$O &$1s_{1/2}$ &6.75&2.85&2.22&2.58&2.73\\
$^{17}_\Lambda$O &$1p_{3/2}$ &6.09&2.82&2.95&2.61&2.70\\
$^{16}$O & &5.84&2.79& &2.64& 2.67\\ \hline 
$^{41}_\Lambda$Ca &$1s_{1/2}$ &7.83&3.52&2.55&3.30&3.42\\
$^{41}_\Lambda$Ca &$1p_{3/2}$ &7.57&3.51&3.17&3.32&3.41\\
$^{40}$Ca & &7.36&3.48$^*$& &3.33&3.38\\ \hline
$^{49}_\Lambda$Ca &$1s_{1/2}$ &7.75&3.54&2.61&3.63&3.45\\
$^{49}_\Lambda$Ca &$1p_{3/2}$ &7.53&3.54&3.24&3.64&3.44\\
$^{48}$Ca & &7.27&3.52& &3.66&3.42\\ \hline
$^{91}_\Lambda$Zr &$1s_{1/2}$ &8.04&4.29&3.02&4.29&4.21\\
$^{91}_\Lambda$Zr &$1p_{3/2}$ &7.98&4.28&3.66&4.30&4.21\\
$^{90}$Zr & &7.79&4.27& &4.31&4.19\\ \hline
$^{209}_\Lambda$Pb &$1s_{1/2}$ &7.39&5.50&3.79&5.67&5.43\\
$^{209}_\Lambda$Pb &$1p_{3/2}$ &7.37&5.49&4.50&5.67&5.43\\
$^{208}$Pb & &7.25&5.49& &5.68&5.43\\
\end{tabular}
\end{center}
\end{table}
%
%

Regarding the effects of the $\Lambda$ on the core nucleons, 
we show also in Fig.~\ref{spectra} the core-nucleon single-particle 
energies for $^{40}$Ca, and $^{41}_\Lambda$Ca for a  
$1s_{1/2}$ $\Lambda$ state. It is interesting to see that the effects 
of the $\Lambda$ work differently on the protons and neutrons in 
$^{41}_\Lambda$Ca. The existence of the $\Lambda$ makes the scalar 
and baryon densities larger, and the scalar and vector potentials 
become stronger. As a consequence, 
the relative strength between the scalar, vector and 
Coulomb potentials change in $^{41}_\Lambda$Ca, 
compared to those in $^{40}$Ca. From 
the calculated results for the proton rms radius,  
$r_p$, and nucleon energy shift shown in Fig.~\ref{spectra},
we can conclude that the Coulomb potential increases more than the gain in 
the difference between the scalar and vector potentials.
We can also say that the scalar potential increases more 
than the vector potential, 
by observing the neutron single-particle energies and the rms radius of the
neutron, $r_n$. (See also Table~\ref{rms}.)

Finally, we show the scalar and vector potential strength 
for $^{17}_\Lambda$O, $^{41}_\Lambda$Ca and 
$^{209}_\Lambda$Pb in Fig.~\ref{potential}.
The difference between the scalar and vector potentials near the 
center of the hypernuclei is typically $\sim 35 - 40$ MeV.
This value is slightly ($\sim 5 - 10$ MeV) larger than that calculated by 
Ma et al.~\cite{zma}, which seems to be the specific origin of the
overbinding in the present calculation. 
Although our calculations are based on 
quark degrees of freedom, in the effective hadronic model 
the whole dynamics of the quarks and gluons 
are absorbed into the parameters and coupling constants appearing 
at the hadron level. In this sense, the effect of the Pauli principle 
at the quark level~\cite{hun,tak}, is not
included between the quarks in the $\Lambda$ and the nucleons.
It seems necessary to incorporate the Pauli principle 
at the quark level, or some 
equivalent effect which would produce a repulsive core in hypernuclear 
systems, if one is to reproduce the experimental binding energies 
in the QMC model.

In summary, we have reported the first results for $\Lambda$ hypernuclei
calculated with the QMC model extended to the flavor SU(3).
The very small spin-orbit force for $\Lambda$ in hypernuclei 
was achieved.
This is a very promising feature and a direct 
consequence of the SU(6) quark model wave function and the quark structure 
of the $\Lambda$ in the QMC model.
However, the calculated single-particle energies 
for the $\Lambda$ tend to be overestimated in comparison with 
the experimental data.
In order to overcome this overbinding problem, 
it may be necessary to introduce a repulsive core  
due to the Pauli principle at the quark level, or some 
equivalent effect. This is a challenging problem for future work.
\vspace{0.5cm}

We would like to thank R. Brockmann for helpful discussions concerning 
the calculation, and P.A.M. Guichon for useful comments.
This work was supported by the Australian Research Council, and 
A.W.T and K.S. acknowledge support from the Japan Society for 
the promotion of Science.
%
%

%
\newpage
%
%
%
%
\setcounter{figure}{0}
%
\begin{figure}[hbt]
\begin{center}
\epsfig{file=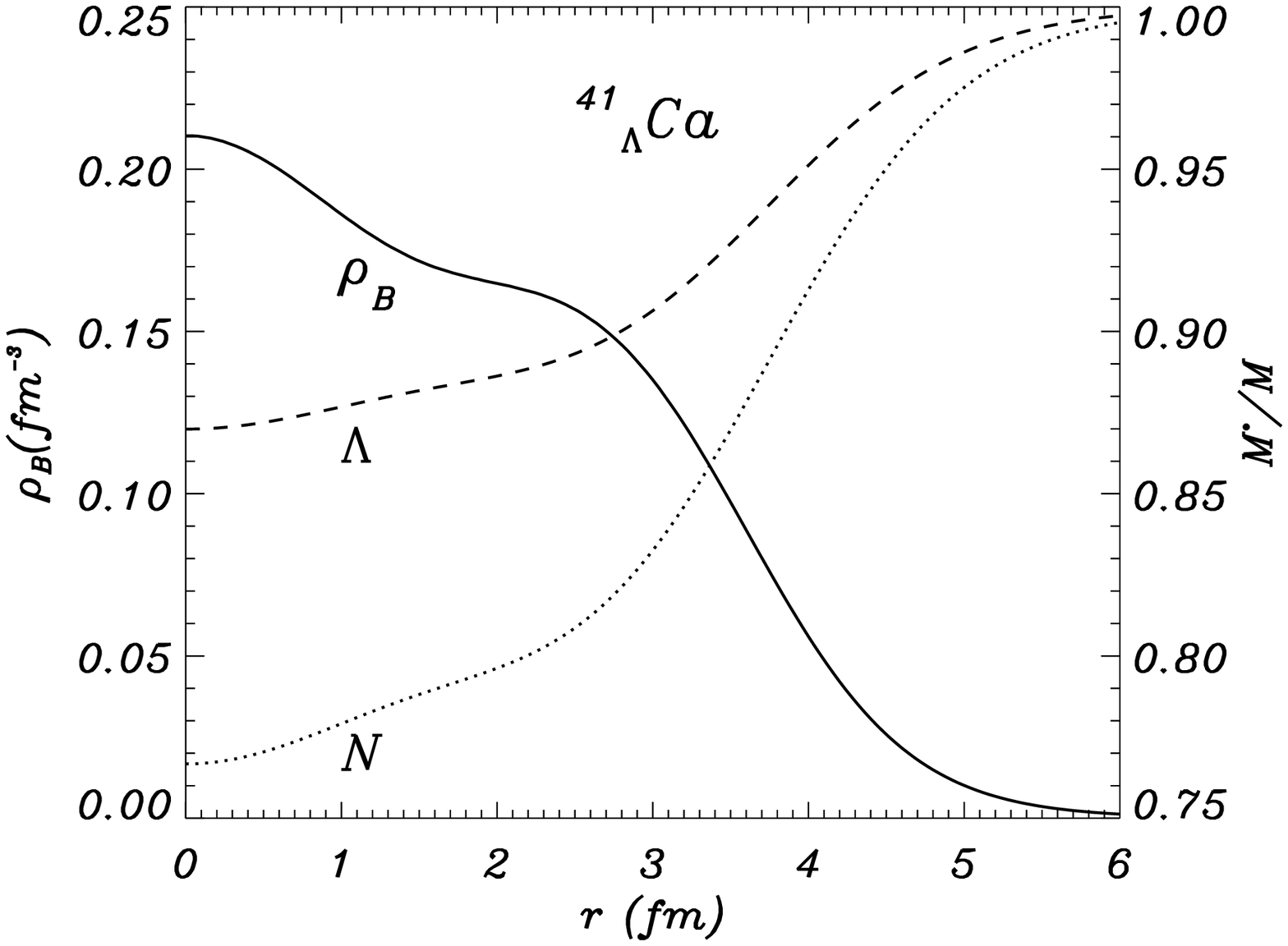,height=7cm}
\caption{(b)}
\label{ymassca}
\end{center}
\end{figure}
%
%
\setcounter{figure}{0}
%
\begin{figure}[hbt]
\begin{center}
\epsfig{file=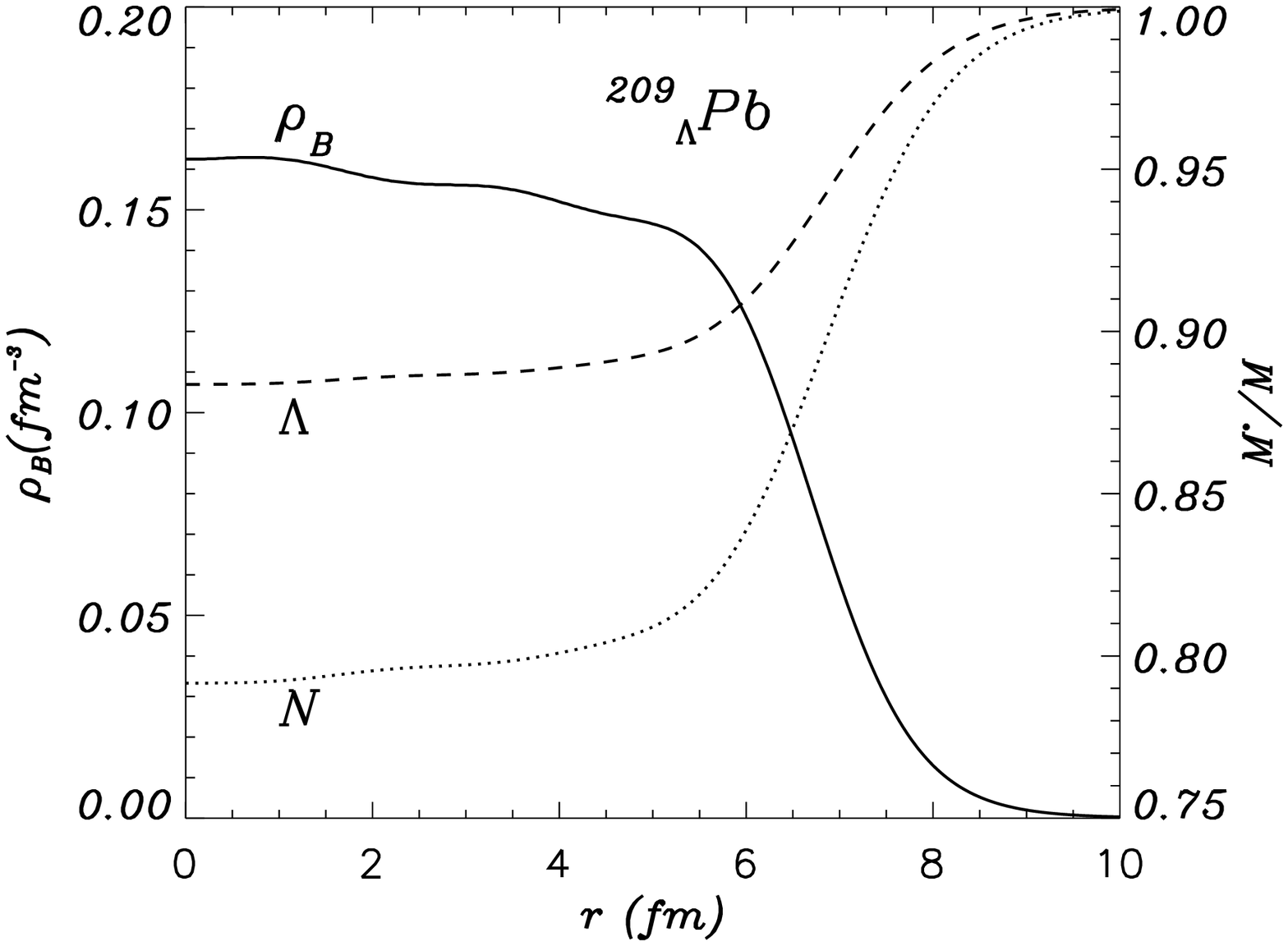,height=7cm}
\caption{(c)}
\setcounter{figure}{0}
\vspace{2cm}
\caption{\newline
Calculated baryon densities, $\rho_B$, 
and effective masses of the nucleon denoted by, N,
and the $\Lambda$ hyperon denoted by, $\Lambda$, in hypernuclei for\quad
(a): $^{17}_\Lambda$O,\quad (b): $^{41}_\Lambda$Ca\quad 
and\quad (c): $^{209}_\Lambda$Pb. All cases are for the 
$1s_{1/2}$ $\Lambda$ state.}
\label{ymasspb}
\end{center}
\end{figure}
%
%
\begin{figure}[hbt]
\begin{center}
\epsfig{file=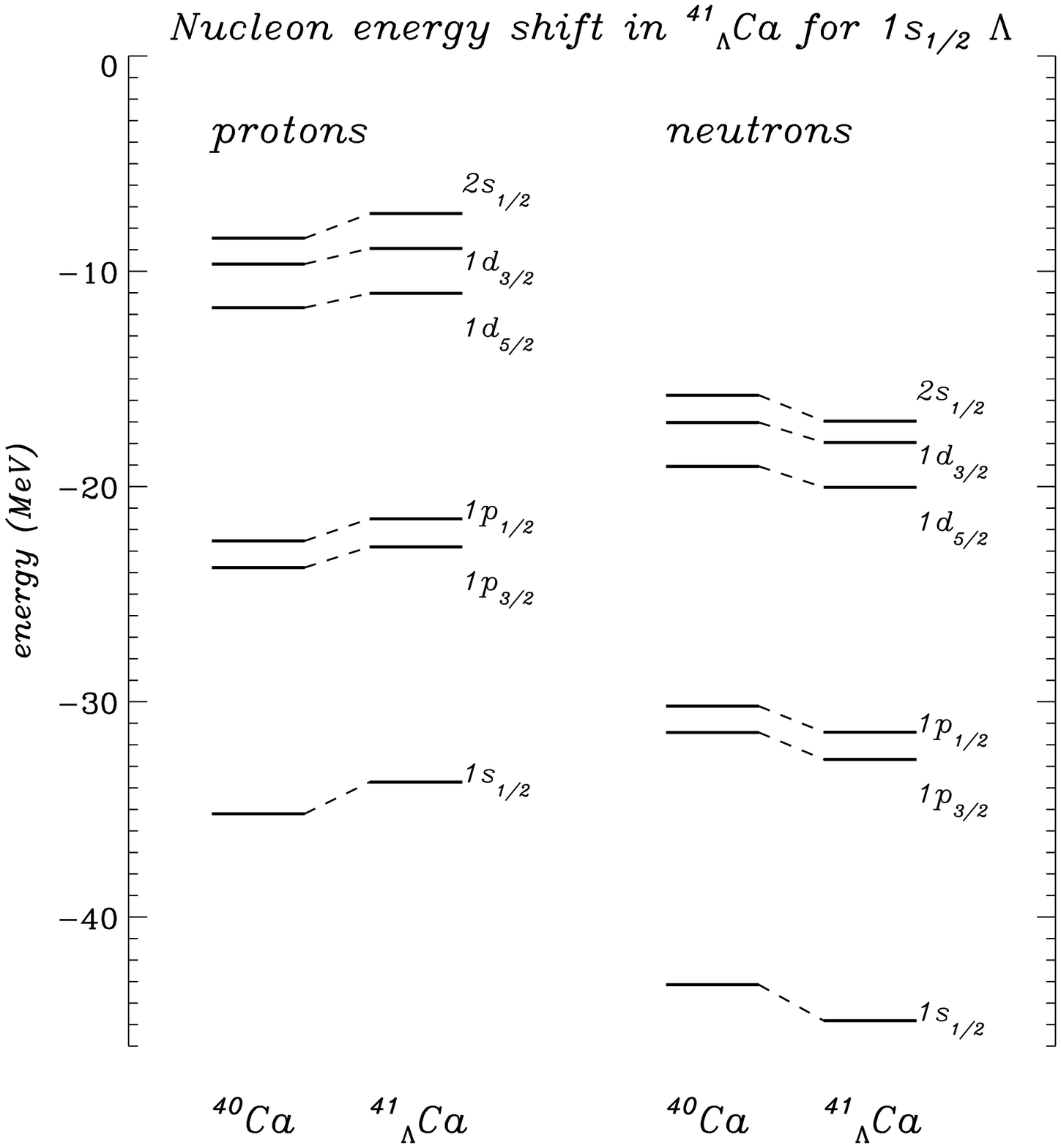,height=14cm}
\caption{Nucleon single-particle energies for $^{40}$Ca, and 
$^{41}_\Lambda$Ca for the $1s_{1/2}$ $\Lambda$ state.}
\label{spectra}
\end{center}
\end{figure}
%
%
\begin{figure}[hbt]
\begin{center}
\epsfig{file=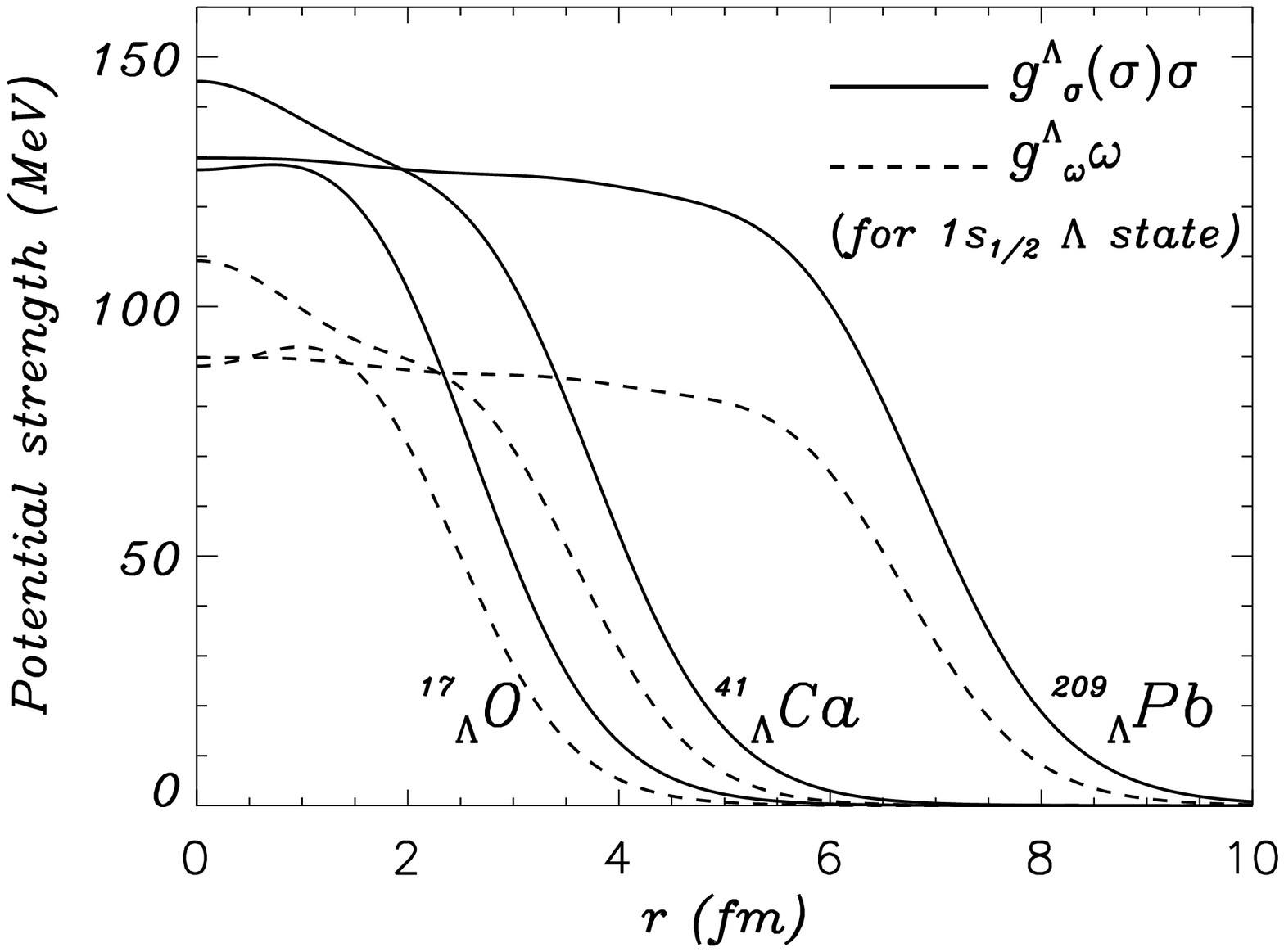,height=9cm}
\caption{Calculated scalar and vector potential strength 
for $^{17}_\Lambda$O, $^{41}_\Lambda$Ca and $^{209}_\Lambda$Pb.}
\label{potential}
\end{center}
\end{figure}
%
%

\begin{thebibliography}{99}
%
\bibitem{finite0}P.A.M. Guichon, K. Saito, E. Rodionov and A.W. Thomas,
Nucl. Phys. {\bf A601} (1996) 349;\\
P.A.M. Guichon, K. Saito and A.W. Thomas, 
Australian Journal of Physics {\bf 50} (1997) 115. 
%
\bibitem{finite1}K. Saito, K. Tsushima and A.W. Thomas, 
Nucl. Phys. {\bf A609} (1996) 339;\\
K. Tsushima, K. Saito and A.W. Thomas, nucl-th/9608062,
to be published in proceedings of the Int. National Symposium on
{\it Non-Nucleonic Degrees of Freedom Detected in Nucleus}, 
Sep. 2-5, 1996, Osaka, Japan.
%
\bibitem{gui}P.A.M. Guichon, Phys. Lett. {\bf B200} (1988) 235.
%
\bibitem{matter}K. Saito and A.W. Thomas, Phys. Lett. {\bf B327} (1994) 9;\\
K. Saito and A.W. Thomas, Phys. Rev. C {\bf 52} (1995) 2789. 
%
\bibitem{blu}P.G. Blunden and G.A. Miller,
Phys. Rev. C {\bf 54} (1996) 359.
%
\bibitem{jin}X. Jin and B.K. Jennings, Phys. Lett. {\bf B374} (1996) 13;\\
X. Jin and B.K. Jennings, Phys. Rev. C {\bf 54} (1996) 1427.
%
\bibitem{qhd}J.D. Walecka, Ann. Phys. (N.Y.) {\bf 83} (1974) 491; \\
B.D. Serot and J.D. Walecka, Adv. Nucl. Phys. {\bf 16} (1986) 1. 
%
\bibitem{finite2}K. Saito and A.W. Thomas, Phys. Rev. C {\bf 51} (1995) 2757;\\
K. Saito, K. Tsushima and A.W. Thomas, Phys. Rev. C {\bf 55} (1997) 2637;\\
K. Saito, K. Tsushima and A.W. Thomas, 
Phys. Rev. C {\bf 56} (1997) 566;\\
K. Saito, K. Tsushima and A.W. Thomas, nucl-th/9608060, 
to be published in proceedings of the Int. National Symposium on
{\it Non-Nucleonic Degrees of Freedom Detected in Nucleus},
Sep. 2-5, 1996, Osaka, Japan.
%
\bibitem{hyper}K. Tsushima, K. Saito J. Haidenbauer 
and A.W. Thomas, ADP-97-26/T261, nucl-th/9707022.
%
\bibitem{bou}A. Bouyssy and J. H\"{u}fner, Phys. Lett. {\bf 64B} (1976) 276.
%
\bibitem{bro}R. Brockmann and W. Weise, Phys. Lett. {\bf 69B} (1977) 167;\\
R. Brockmann and W. Weise, Nucl. Phys. {\bf A355} (1981) 365.
%
\bibitem{bru}W. Br\"{u}ckner et al., Phys. Lett. {\bf 79B} (1978) 157.
%
\bibitem{pir}H.J. Pirner, Phys. Lett. {\bf 85B} (1979) 190.
%
\bibitem{nob}J.V. Noble, Phys. Lett. {\bf 89B} (1980) 325.
%
\bibitem{dov}C.B. Dover and A. Gal, Prog. Part. Nucl. Phys. {\bf 12} 
 (1984) 171.
%
\bibitem{jen}B.K. Jennings, Phys. Lett. {\bf 246B} (1990) 325.
%
\bibitem{bog}J. Boguta and S. Bohrmann, Phys. Lett. {\bf 102B} (1981) 93.
%
\bibitem{bou2}A. Bouyssy, Nucl. Phys. {\bf A381} (1982) 445.
%
\bibitem{ruf}M. Rufa, H. St\"{o}cker, P-G Reinhard, 
J. Maruhn and W. Greiner, J. Phys. {\bf G13} (1987) L143.
%
\bibitem{mar}J. Mare\v s and J. \v Zofka, Z. Phys. {\bf A333} (1989) 209.
%
\bibitem{coh}Joseph Cohen and H. J. Weber, Phys, Rev. C {\bf 44} (1991) 1181.
%
\bibitem{chi}M. Chiapparini, A.O. Gattone and B.K. Jennings, 
Nucl Phys. {\bf A529} (1991) 589;\\
E.D. Cooper, B.K. Jennings, J. Mare\v s, Nucl. Phys. {\bf A580} (1994) 419;\\
J. Mare\v s and B.K. Jennings, Phys. Rev. C {\bf 49} (1994) 2472;\\
J. Mare\v s, B.K. Jennings and E.D. Cooper, Prog. Theor. Phys. Supp.
{\bf 117} (1994) 415;\\
J. Mare\v s and B.K. Jennings, Nucl. Phys. {\bf A585} (1995) 347c.
%
\bibitem{gle}N.K. Glendenning, D. Von-Eiff, M. Haft, 
H. Lenske and M.K. Weigel, Phys. Rev. C {\bf 48} (1993) 889.
%
\bibitem{sug}Y. Sugahara and H. Toki, Prog. Theor. Phys. 
(1994) 803.
%
\bibitem{ine}F. Ineichen, D. Von-Eiff and M.K. Weigel, 
J. Phys. {\bf G22} (1996) 1421. 
%
\bibitem{zma}Zhong-yu Ma, J. Speth, S. Krewald, Bao-qiu Chen and A. Reuber, 
Nucl. Phys. {\bf A608} (1996) 305.
%
\bibitem{pro}Prog. Theor. Phys. Supplement, {\bf 117} (1994),
edited by T. Motoba, Y. Akaishi and K. Ikeda, and references therein. 
%
\bibitem{pov}B. Povh, Nucl. Phys. {\bf A335} (1980) 233.
%
\bibitem{chr}R.E. Chrien {\bf A478} (1988) 705c.
%
\bibitem{mitbag}T. DeGrand, R.L. Jaffe, K. Johnson, and J. Kiskis, 
Phys. Rev. D {\bf 12} (1975) 2060.
%
\bibitem{pil}P.H. Pile et al., Phys. Rev. Lett. {\bf 66} (1991) 2585.
%
\bibitem{aji}S. Ajimura et al., Nucl. Phys. {\bf A585} (1995) 173c.
%
\bibitem{bod}A.R. Bodmer, Phys. Rev. {\bf 141} (1966) 1387;\\
R.H. Dalitz, R.C. Herndon, Y.C. Tang, Nucl. Phys. {\bf B47} (1972) 109.
%
\bibitem{hun}E.V. Hungerford and L.C. Biedenharn,
Phys. Lett. {\bf 142B} (1984) 232. 
%
\bibitem{tak}S. Takeuchi and K. Shimizu, Phys. Lett. {\bf 179B} (1986) 197;\\
M. Oka, K. Shimizu and K. Yazaki, Nucl. Phys. {\bf A464} (1987) 700;\\
K. Shimizu, Nucl. Phys. {\bf A547} (1992) 265c.
%
\bibitem{str}U. Straub et al., Nucl. Phys. {\bf 483} (1988) 686.
%
\bibitem{fur}R.J. Furnstahl and B.D. Serot, Nucl. Phys. {\bf A468} (1987) 539.
%
\bibitem{coh}J. Cohen, Phys. Rev. C {\bf 48} (1993) 1346.
%
\bibitem{coh2}J. Cohen and R.J. Furnstahl, Phys. Rev. C {\bf 35} (1987) 2231.
%
\bibitem{neg}J.W. Negele, Phys. Rev. {\bf 4} (1970) 1260.
%
\end{thebibliography}
\end{document}